\documentclass[preprint,aps,pra,showpacs,floatfix]{revtex4-1}
\usepackage[utf8]{inputenc}
\usepackage[OT1]{fontenc}
\usepackage{amsmath}
\usepackage{amsfonts}
\usepackage{amssymb}
\usepackage{bm}
\usepackage{graphicx}
\usepackage[left=2cm,right=2cm,top=2cm,bottom=2cm]{geometry}
\usepackage{longtable}

\begin{document}

\title{Ionization energies along beryllium isoelectronic sequence}

\author{A. V. Malyshev$^{1,2}$, A. V. Volotka$^{1,3}$, D. A. Glazov$^{1,2,3}$, I. I. Tupitsyn$^{1}$,  V. M. Shabaev$^{1}$, and G. Plunien$^{3}$}

\affiliation{
$^1$ Department of Physics, St. Petersburg State University,
Ulianovskaya 1, Petrodvorets, 198504 St. Petersburg, Russia \\
$^2$ State Scientific Centre ``Institute for Theoretical and Experimental Physics'' of National Research Centre ``Kurchatov Institute'', B. Cheremushkinskaya st. 25, 117218 Moscow, Russia \\
$^3$ Institut f\"ur Theoretische Physik, Technische Universit\"at Dresden,
Mommsenstra{\ss}e 13, D-01062 Dresden, Germany \\
}

\begin{abstract}
Ionization energies for the ground state of berylliumlike ions with nuclear charge numbers in the range $16 \leqslant Z \leqslant 96$ are rigorously evaluated. The calculations merge the \textit{ab initio} QED treatment in the first and second orders of the perturbation theory in the fine-structure constant $\alpha$ with the third- and higher-order electron-correlation contributions evaluated within the Breit approximation. The nuclear recoil and nuclear polarization effects are taken into account. The accuracy of the ionization energies obtained has been significantly improved  in comparison with previous calculations. 
\end{abstract}

\maketitle

\section{Introduction \label{sec:0}}

During the past four decades, starting from the pioneering works on calculations of one-electron self-energy \cite{Mohr:1974:26:note} and vacuum polarization \cite{Soff:1988:5066,Manakov:1989:1167:note} corrections, considerable progress has been achieved in the theoretical understanding of  the calculations of quantum electrodynamics (QED) corrections to all orders in $\alpha Z$, where $\alpha$ is the fine-structure constant and $Z$ is the nuclear charge number. To date, state-of-the-art QED calculations include all corrections up to the second order in $\alpha$, for review see Refs.~\cite{Sapirstein:2008:25,Shabaev:2008:1220,Shabaev:2011:60,Glazov:2011:71,Volotka:2013:636} and references therein. High-precision experiments to measure the binding and transition energies in highly charged ions \cite{Schweppe:1991:1434,Stoehlker:1993:2184,Beiersdorfer:1998:1944,Beiersdorfer:1998:3022,
Bosselmann:1999:1874,Stoehlker:2000:3109,Brandau:2003:073202,Draganic:2003:183001,
Gumberidze:2004:203004,Gumberidze:2005:223001,Beiersdorfer:2005:233003,Mackel:2011:143002,
Kubicek:2014:032508,Bernhardt:2015:144008} are sensitive to the second-order QED corrections. They confirm theoretical predictions to a high accuracy, and thereby provide a unique opportunity for tests of QED in the strong Coulomb field.

Among rigorous QED calculations of the energy levels of highly charged ions, see \cite{Persson:1996:204,Yerokhin:1997:361,Mohr:1998:227,Yerokhin:2001:032109,Yerokhin:2003:203,
Artemyev:2007:173004,Yerokhin:2008:062510,Kozhedub:2010:042513,Artemyev:2013:032518,
Johnson:1985:405,Drake:1988:586,Artemyev:2005:062104,Sapirstein:2011:012504} and references therein, a particular place is occupied with the works where the ionization energies are investigated. The ionization energies allow one to relate masses of ions with different number of electrons, and hence are of great interest for the tasks of mass spectrometry  \cite{Repp:2012:983,Myers:2013:107}.
Up to date, the high-precision \textit{ab initio} QED calculations of the ionization energies have been performed for hydrogen \cite{Johnson:1985:405,Yerokhin:hydrogen:preprint}, helium \cite{Drake:1988:586,Artemyev:2005:062104}, and lithium \cite{Sapirstein:2011:012504} isoelectronic sequences. There are many nonrelativistic and relativistic calculations of the ground-state energies of berylliumlike ions in the literature \cite{Chung:1993:1740,Biemont:1999:117,Chaudhuri:2000:5129,Rodrigues:2004:117,Gu:2005:267,
Huang:2006:23,Kramida:2006:457,Yong-Qiang:2008:3627,Argaman:2013:042504,
Pathak:2014:042510,Yerokhin:2014:022509}. Some of these works somehow include the radiative and nuclear recoil corrections. As a rule, only the first-order QED effects are included, the many-electron QED effects are treated  with the use of one-electron  approximations or semiempirical methods. The necessity of a more rigorous QED treatment to improve  the theoretical accuracy has been pointed out in Ref. \cite{Yerokhin:2014:022509}.

Recently, in Ref.~\cite{Malyshev:2014:062517}, we have calculated the ground-state binding energies for the even-$Z$ berylliumlike ions in the range $18 \leqslant Z \leqslant 96$. We have also evaluated the  ionization energies for these ions by subtracting the binding energies of corresponding Li-like ions taken from the literature. The main goal of the present investigation is to calculate  the ground-state  ionization potentials for these ions  directly and self-consistently including the QED corrections up to the second order in $\alpha$ and the electron-correlation effects to all orders in $1/Z$. We also extend these calculations to all other ions along the beryllium isoelectronic sequence  with nuclear charge $16 \leqslant Z \leqslant 96$. 
 
The paper is organized as follows.  The procedure of calculation of the ionization energies is described in Sec.~\ref{sec:1}. In Sec.~\ref{sec:2} we present our numerical results and compare them with the previous computations. Section \ref{sec:3} is reserved for a brief summary.

The relativistic units ($\hbar = c = 1$) and the Heaviside charge unit ($\alpha = e^2/4\pi, e<0$) are used throughout the paper.

\section{Basic formulas \label{sec:1}}

The Furry picture is the standard starting point for investigation of the properties of heavy few-electron ions within QED.
In this picture interaction of electrons with the Coulomb field of the nucleus $V_{\rm{nucl}}$ is taken into account to all orders in $\alpha Z$ from the very beginning:
\begin{equation}
\left[-i \bm {\alpha} \cdot \nabla + \beta m + V_{\rm{nucl}}( \bm{r} )\right] \psi_n (\bm{r}) = \varepsilon_n \psi_n (\bm{r}).
\label{DirEq}
\end{equation}
The interelectronic interaction and QED effects have to be accounted for 
by perturbation theory. A convenient approach to construct the QED perturbation series is the two-time Green function (TTGF) method \cite{TTGF}. 

Naturally, such an approach works well when the interelectronic  interaction is small in comparison with the binding energies of all electrons of the system. Otherwise, the convergence of the perturbation series is often slow. It is possible to accelerate the convergence of the perturbation theory by transition to the extended Furry picture. This implies the  replacement of the nuclear potential in Eq. (\ref{DirEq}) with an effective potential:
\begin{equation}
V_{\rm{nucl}}( \bm{r} ) \rightarrow V_{\rm{eff}}( \bm{r} ) = V_{\rm{nucl}}( \bm{r} ) + V_{\rm{scr}}( \bm{r} ).
\label{EffPot}
\end{equation} 
The local screening potential $V_{\rm{scr}}( \bm{r} )$ in Eq. (\ref{EffPot}) models the screening of the nuclear potential by the other electrons  partly accounting for the interelectronic interaction in the zeroth-order Hamiltonian.  The interaction with the potential $\delta V( \bm{r} ) = -V_{\rm{scr}}( \bm{r} )$ is to be accounted for perturbatively to avoid the double counting of the screening effects. Furthermore, the calculation of the $1s^22s^2$ state energy starting from the pure Coulomb field as the zeroth order approximation is difficult, because of the quasidegeneracy with the $1s^2(2p_{1/2})^2$ state. The application of some screening potentials can remove this degeneracy and simplify calculations significantly.
The extended Furry picture was successfully employed in QED calculations of the energy levels \cite{Sapirstein:2001:022502,Sapirstein:2002:042501,Chen:2006:042510,Artemyev:2007:173004,
Yerokhin:2007:062501,Kozhedub:2010:042513,Sapirstein:2011:012504,Artemyev:2013:032518}, the hyperfine splitting \cite{Sapirstein:2001:032506,Sapirstein:2003:022512,Sapirstein:2006:042513,Oreshkina:2007:889:note,
Kozhedub:2007:012511,Volotka:2008:062507}, and the $g$-factor \cite{Glazov:2006:330,Glazov:2013:014014,Volotka:2014:253004}. 

In this paper we apply four different types of the screening potential.  All these potentials are spherically symmetric by construction.  Our first and simplest choice is the core-Hartree (CH) potential induced by the $1s^2$ closed shell. It can be obtained from the radial charge density of two $1s$ electrons:
\begin{gather}
V_{\rm CH}(r) = \alpha \int_0^\infty d r' \frac{1}{r_>} \rho_{\rm CH} (r'),  \label{CHpot} \\
\rho_{\rm CH}(r) = 2 \left[ G_{1s}^2(r) + F_{1s}^2(r)  \right], \qquad 
\int_0^\infty \rho_{\rm CH}(r) dr = 2,    \label{CHdens} 
\end{gather}
here $G/r$ and $F/r$ are large and small radial components of the Dirac wave function. 

Three other types of the screening potential were constructed for both three-electron  ($1s^2 2s$) and four-electron ($1s^2 2s^2$)  configurations, so that the total number of the screening potentials employed is equal to seven. In what follows we label these potentials with the indices 3 and 4, depending on the electronic configuration used. 
The first type of the screening potential generated for both configurations is the local Dirac-Fock (LDF) potential \cite{pot:LDF}. This potential is constructed by inversion of the radial Dirac equation with the radial wave functions obtained in the Dirac-Fock approximation. 
Two other screening potentials arise from the density-functional theory. 
In terms of the total radial charge density $\rho_{ t}$ of all electrons in the configuration the Kohn-Sham (KS) potential reads as follows:
\begin{gather}
V_{\rm{KS}}(r) = \alpha \int_0^\infty d r' \frac{1}{r_>} \rho_t(r') - \frac{2}{3} \frac{\alpha}{r} \left( \frac{81}{32\pi^2}  r \rho_t(r) \right)^{1/3},\label{KSpot} \\
\rho_t(r) = \rho_{\rm CH}(r) + \left(N-2\right) \left[ G_{2s}^2(r) + F_{2s}^2(r) \right], \qquad \int_0^\infty \rho_{t}(r) dr = N,    \label{KSdens} 
\end{gather} 
where $N=3$ or 4 depending on the electron configuration.  We introduce the  Latter correction \cite{pot:Latter} to improve the asymptotic behavior of the KS potentials at large $r$.
Finally, the last type of the screening potential generated for  $1s^2 2s$ and $1s^2 2s^2$ configurations is the Perdew-Zunger (PZ) potential \cite{pot:PZ}. 
Summarizing the description of the screening potentials used in the present work it is worth noting that they have different asymptotic behavior. The potentials $V_{\rm CH}$, $V_{\rm KS3}$, $V_{\rm LDF3}$ and $V_{\rm PZ3}$ behave like $2\alpha/r$ at large distances  ${r}$, whereas  the potentials $V_{\rm KS4}$, $V_{\rm LDF4}$, $V_{\rm PZ4}$  behave like $3\alpha/r$. In other words, if we set    $N_{\rm scr}=N-1$ for the LDF, KS and PZ potentials, and $N_{\rm scr}=2$ for the CH potential, then all screening potentials behave like $N_{\rm scr}\alpha/r$. In this case $eN_{\rm scr}$ has the meaning of the total charge of the screening cloud.

The calculation of the ionization potentials of berylliumlike ions in the present work follows in general the scheme which was used in Ref. \cite{Malyshev:2014:062517} for the evaluation of the binding energies. Nevertheless, there are some modifications in the calculation procedure. Below, we briefly discuss the main stages of the calculations.  

The evaluation of the ionization potentials can be conveniently divided into several steps.
At first stage it is necessary to solve Eq. (\ref{DirEq}) for the state under consideration with an effective potential (\ref{EffPot}). Moreover, in the bound-state QED calculations one needs to have a quasicomplete basis set of the Dirac equation solutions for
representation of the Dirac-Coulomb Green function
\begin{equation}
G(\omega, \bm{r}_1,\bm{r}_2) = \sum_n \frac{\psi_n(\bm{r}_1)\psi^\dagger_n(\bm{r}_2)}{\omega - \varepsilon_n(1-i0)}.
\label{GreenF}
\end{equation}
The set of one-electron wave functions was evaluated employing the dual kinetic balance (DKB) method  \cite{splines:DKB} with the basis functions constructed from the B-splines \cite{splines:B}.

\begin{figure}
\begin{center}
\includegraphics[width=12cm]{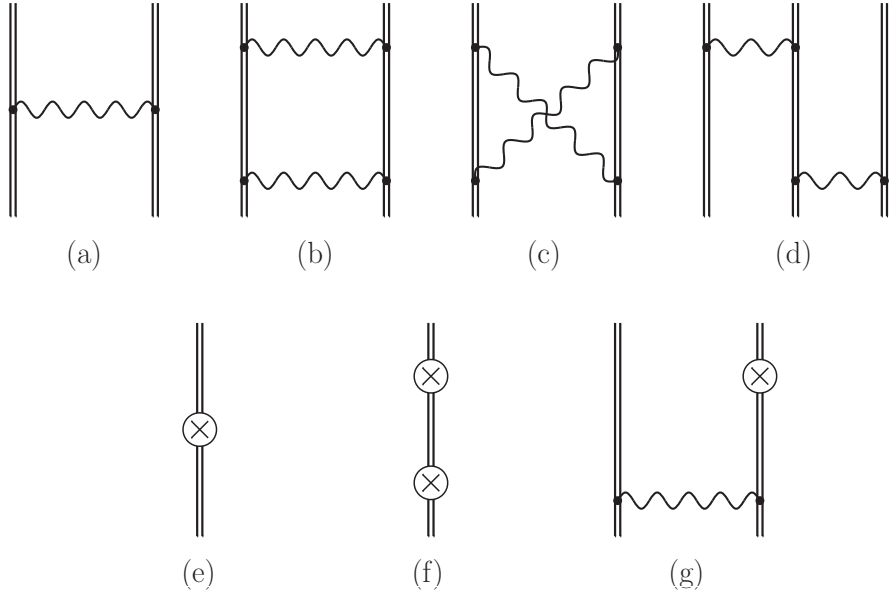}
\caption{\label{fig:int}
The interelectronic-interaction diagrams. 
The double line denotes the  electron propagator in the effective potential (\ref{EffPot}). 
The wavy line corresponds to the photon propagator.
The symbol $\otimes$ represents the local screening potential counterterm.}
\end{center}
\end{figure}

The application of the screening potentials allows one  to partly  take  into account  the interaction between electrons. We have to consider the remaining interelectronic interaction by perturbation theory. In  Fig.~\ref{fig:int} the corresponding set of  first- and second-order  Feynman diagrams  is depicted. The double line represents the electron propagator in the effective potential. The circle with a cross corresponds to the screening potential counterterm. The ionization energy of the $2s$ electron can be obtained by subtracting the binding energy of the Li-like ion from the binding energy of the Be-like ion. Diagrams containing only the interaction  between the $1s$ electrons contribute to both binding energies and cancel each other in the difference. For this reason, here and further throughout the paper we deal only with those diagrams, where one or two of the incoming (outgoing) electron lines  correspond to the $2s$ electrons. The calculation formulas for the diagrams (a)-(d) in Fig.~\ref{fig:int} can be found, e.g., in Refs.~\cite{Shabaev:1994:4489,Yerokhin:2001:032109}. As it was noted in Ref.~\cite{Malyshev:2014:062517}, only a minor modification of the standard procedures is necessary to perform the calculation of the $2s^2$ interaction. 
The formulas for the corrections corresponding to the diagrams (e)-(g) can be easily obtained with the use of the TTGF method:
\begin{eqnarray}
\Delta E_e &=&   V_{aa}, \label{contr_e} \\
\Delta E_f &=& \sum_{n \neq a} \frac{|V_{an}|^2}{\varepsilon_{a} - \varepsilon_n},   \label{contr_f} \\
\Delta E_g &=& \left. 4 \sum_{n \neq a} 
\frac{I_{a \bar{a}; n \bar{a}} V_{na} }{\varepsilon_a - \varepsilon_n} \right|_{\mu_{\bar{a}}=-\mu_a} \nonumber \\
&+& \sum_{\mu_b} \Bigg\{  2  \Bigg[ 
\sum_{n \neq b} \frac{I_{b a ; n a} V_{nb} }{\varepsilon_b - \varepsilon_n} 
 + \sum_{n \neq a} \frac{I_{a b ; n b} V_{na} }{\varepsilon_a - \varepsilon_n} \Bigg]   \nonumber\\
&+& \left( V_{aa} - V_{bb} \right) I'_{baab}(\varepsilon_b - \varepsilon_a)\Bigg\},  
\label{contr_g}
\end{eqnarray} 
where $a=2s$, $b=1s$, $V_{ab}=\langle a | \delta V | b \rangle = \langle a | -V_{\rm{scr}} | b \rangle$, $I_{abcd}(\omega) = \langle ab| I(\omega)|cd\rangle$, $I(\omega)=e^2 \alpha^\mu \alpha^\nu D_{\mu\nu}(\omega)$, $D_{\mu\nu}(\omega)$ is the photon propagator, $I'_{abcd}(\omega)=\langle ab| \tfrac{\partial}{\partial \omega}I(\omega)|cd\rangle$, $I_{ab;cd}=\langle ab| I(\Delta_{bd})|cd\rangle - \langle ba| I(\Delta_{ad})|cd\rangle$, and 
$\Delta_{ab} = \varepsilon_a - \varepsilon_b$.

Next, we have to take into account the interelectronic-interaction corrections of the third and higher orders.
We have evaluated these contributions within the Breit approximation. At first, we have calculated the binding energies of the $1s^22s$ and $1s^2 2s^2$ states solving the Dirac-Coulomb-Breit equation by means of the configuration-interaction Dirac-Fock-Sturm method (CI-DFS) \cite{CI,CI2}. Then, the contributions of third and higher orders   to these binding energies, $E_{1s^2 2s}^{(\geqslant 3)}$ and $E_{1s^2 2s^2}^{(\geqslant 3)}$,  were extracted  from the total CI-DFS energies  using the numerical procedure described, e.g., in Refs. \cite{Kozhedub:2010:042513,Artemyev:2013:032518}. 
Finally,  we took the difference $E_{1s^2 2s^2}^{(\geqslant 3)}-E_{1s^2 2s}^{(\geqslant 3)}$ to obtain the desired contribution $E_{\rm int,Breit}^{(\geqslant 3)}$  to the ionization potential of the $2s$ electron. To estimate the accuracy of our numerical procedure we have compared the first- and second-order interelectronic-interaction contributions to the ionization potential, $E_{\rm int,Breit}^{(1)}$ and $E_{\rm int,Breit}^{(2)}$, calculated within CI-DFS method with analogous quantities evaluated  with the use of our code for the QED calculations.  For this reason the diagrams (a), (b), (d)-(g)  in Fig.~\ref{fig:int} have been evaluated in the Coulomb gauge neglecting the contribution of the negative-energy continuum and energy dependence of the interelectronic-interaction operator.  We have found a very good agreement between those two independent approaches for all types of the screening potentials.

\begin{figure}
\begin{center}
\includegraphics[width=13cm]{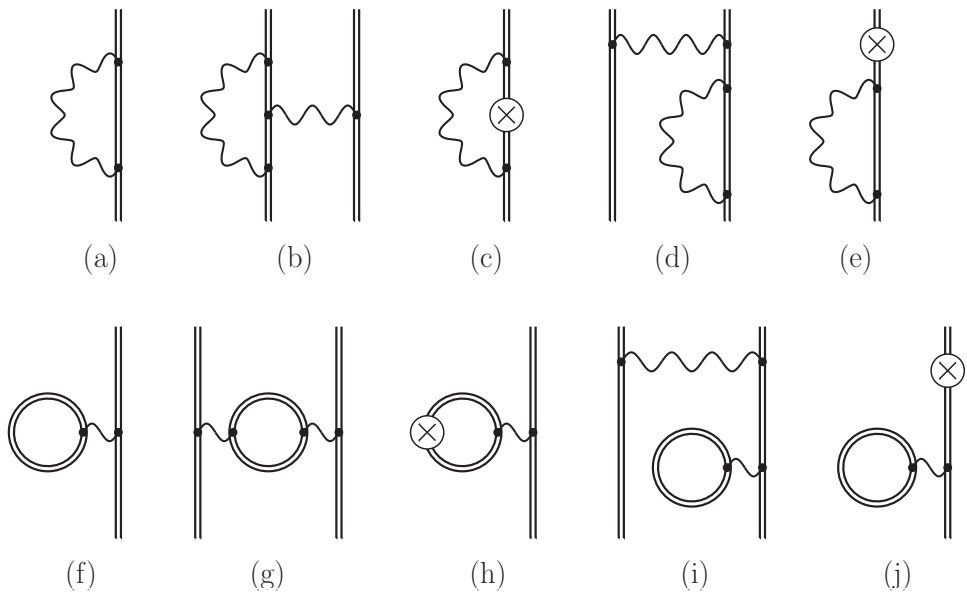}
\caption{\label{fig:se}
First- and second-order QED diagrams (excluding the one-electron two-loop diagrams). The notations are the same as in Fig. \ref{fig:int}.}
\end{center}
\end{figure}

Next, one should consider the contribution of the radiative corrections. Figure \ref{fig:se} displays  all relevant first- and second-order QED diagrams with the exception of one-electron two-loop graphs, which we will discuss below.  The diagrams from the first line are known as the self-energy (SE) diagrams, while the diagrams in the second line are referred to as the vacuum polarization (VP) diagrams. The formal expressions for these contributions can be easily obtained within the TTGF method, they are  presented, e.g., in Ref.~ \cite{Kozhedub:2010:042513}. The formal expressions suffer from ultraviolet  divergences. The way for cancellation of the divergences in the SE diagrams is discussed in detail in Refs. \cite{Yerokhin:1999:3522,Yerokhin:1999:800},  in the present work we follow the renormalization scheme described there. 

The VP contributions are conveniently divided into the Uehling and Wichmann-Kroll parts.  The renormalized expression for the Uehling potential is well known:
\begin{eqnarray}
V_{\rm Uehl}(r)  &=&  -\alpha Z \frac{2\alpha}{3\pi} \int^\infty_0 dr' \; 4 \pi r' \rho_{\rm nucl}(r')
\int^\infty_1 dt \; \left( 1+\frac{1}{2t^2} \right) \frac{\sqrt{t^2-1}}{t^2}  \nonumber \\
 && \times \frac{ \{ \exp(-2m|r-r'|t)-\exp(-2m(r+r')t) \} }{4mrt}, \label{Uepot} 
\end{eqnarray}
where $\rho_{\rm nucl}(r)$ is the nuclear charge density, normalized to unity.
To account for the screening effect on the Uehling potential in the diagrams (f), (i)-(j), one should replace $Z\rho_{\rm nucl}(r)$ by $Z\rho_{\rm nucl}(r) - N_{\rm scr} \rho_{\rm scr}(r)$, where $\rho_{\rm scr}(r)$ is the charge density of the electron screening cloud, normalized to unity. In order to calculate the contribution of the counterterm diagram (h) one has to replace $Z\rho_{\rm nucl}(r)$ by $N_{\rm scr} \rho_{\rm scr}(r)$ in Eq.~(\ref{Uepot}).
The renormalized Uehling operator which corresponds to the diagram (g) has the form
\begin{eqnarray}
I_{\rm Uehl}(\varepsilon, \bm{r}, \bm{r}') &=&
\alpha \frac{\alpha_{1\mu}\alpha^\mu_2}{|\bm{r}-\bm{r}'|}
\frac{2}{3}\frac{\alpha}{\pi}
\int^\infty_1 dt \; \left( 1+\frac{1}{2t^2} \right) \frac{\sqrt{t^2-1}}{t^2} \nonumber \\
&& \times \exp(-\sqrt{(2mt)^2-\varepsilon^2}|\bm{r}-\bm{r}'|),
\label{UehlOper}
\end{eqnarray}
where $\varepsilon$ is the energy of the transmitted photon.

The Wichmann-Kroll contribution was calculated for the diagrams (f), (i) and (j)  using the expression:
\begin{eqnarray}
V_{\rm WK}(r) &=& \frac{2\alpha}{\pi} \sum_{\kappa=\pm1}^{\pm\infty} |\kappa| \nonumber \\
&& \times \int^\infty_0  d\omega\;  \int^\infty_0 dy \;  y^2 \int^\infty_0 dz \; z^2 \frac{1}{{\rm max}(r,y)}  \nonumber \\
&& \times V_{\rm eff}(z) \sum_{i,k=1}^2 {\rm Re} \{ F^{ik}_\kappa(i\omega,y,z)[G^{ik}_\kappa(i\omega,y,z)-F^{ik}_\kappa(i\omega,y,z)] \},
\label{WKpot}
\end{eqnarray}
where $F^{ik}_\kappa$  and $G^{ik}_\kappa$ are the radial components of the partial contributions to the free- and bound-electron Green function. It turns out that calculation of Eq.~(\ref{WKpot}) using the spectral decomposition (\ref{GreenF}) with the B-spline finite basis set is rather problematic. Therefore, we have evaluated the radial Green functions for all screening potentials numerically, solving a corresponding system of differential equations. The Wichmann-Kroll contribution to the diagram (g) is relatively small \cite{Artemyev:1997:3529,Artemyev:1999:45} and has been neglected, together with the related contribution to the diagram (h).

The consideration of the two-loop one-electron contributions completes the rigorous QED treatment in the second order in $\alpha$. Evaluation of these corrections to all orders in $\alpha Z$ is a very demanding problem which has not yet been solved completely. The most remarkable progress in this area was achieved by Yerokhin \textit{et al.} \cite{Yerokhin:2006:253004,Yerokhin:2008:062510}. In Ref.~\cite{Yerokhin:2006:253004}  the complete gauge-invariant set of the two-loop self-energy diagrams was calculated for high-$Z$ ions. The so-called SEVP, VPVP, and S(VP)E sets were tabulated in  Ref.~\cite{Yerokhin:2008:062510} using the \textit{free-loop approximation} if the rigorous calculation was not possible. 
We take the two-loop one-electron corrections from Refs. \cite{Yerokhin:2006:253004,Yerokhin:2008:062510,Yerokhin:hydrogen:preprint}.

All corrections considered so far correspond to the approximation of an infinite nuclear mass. Nevertheless, high precision calculations of energy levels in highly charged ions must include the nuclear recoil corrections. The full relativistic theory of the recoil effect can be formulated only in the framework of QED. Such a theory  to the first order in $m/M$ ($M$ is the nuclear mass) and to all orders in $\alpha Z$ was worked out in Refs.~\cite{Shabaev:1985:394:note,Shabaev:1988:107:note}  
(see also Refs. \cite{Shabaev:1998:59,Adkins:2007:042508} and references therein). 
Within the Breit approximation  the following many-body Hamiltonian describes the contribution of the recoil effect to the binding energy \cite{Shabaev:1985:394:note,Shabaev:1988:107:note,Palmer:1987:5987}
\begin{equation}
H_{M}=\frac{1}{2M} \sum_{i,j} \left\{ \bm{p}_i \cdot \bm{p}_j  - \frac{\alpha Z}{r_i} \left[ \bm{\alpha}_i + \frac{(\bm{\alpha}_i \cdot \bm{r}_i) \bm{r}_i }{r_i^2} \right] \cdot \bm{p}_j \right\}.
\label{BrRecoil}
\end{equation}
Evaluating the expectation values of the relativistic operator (\ref{BrRecoil}) with the CI-DFS wave functions of $1s^2 2s$ and $1s^2 2s^2$ states and taking the difference we obtain  the nuclear recoil contribution to the ionization potential of the $2s$ electron in the Breit approximation. 

The nuclear recoil effects which are beyond the Breit approximation are termed as the QED recoil effects. In the present work we calculate these corrections in the zeroth order in $1/Z$. To this order only one-electron terms contribute. The two-electron terms vanish, since the $1s$ and $2s$ electrons have the same parity \cite{Artemyev:1995:1884}. For the point-charge nucleus, the QED recoil correction to first order in $m/M$, to zeroth order in $\alpha$, and to all orders in $\alpha Z$  is given by \cite{Shabaev:1985:394:note,Shabaev:1988:107:note}:
\begin{eqnarray}
\Delta E &=& \frac{i}{2\pi M} \int_{-\infty}^{\infty} d \omega \;
\left\langle a \left| \left( {\bm D}(\omega)-\frac{[{\bm p},V]}{\omega+i0} \right) \right. \right. \nonumber \\
&& \times G(\omega + \varepsilon_a) \left. \left. \left( {\bm D}(\omega)+\frac{[{\bm p},V]}{\omega+i0} \right)
\right| a  \right\rangle.
\label{QedReoil}
\end{eqnarray}
Here ${\bm p}$ is the momentum operator, $D_m(\omega)=-4\pi \alpha Z \alpha_l D_{lm}(\omega)$, and $D_{lm}$ is the transverse part of the photon propagator in the Coulomb gauge. The nuclear size corrections to the QED recoil effect can be partly accounted for by replacement of the potential $V$, the Dirac energy  $\varepsilon_a$, the wave function $|a\rangle$, and the Green function $G$ in Eq.~(\ref{QedReoil}) with the corresponding  quantities for the extended nucleus \cite{Shabaev:1998:59}. Such calculations were carried out in Refs.~\cite{Shabaev:1998:4235,Shabaev:1999:493} for $1s$ and $2s$ states of H-like ions. In the present work we have performed the calculations of the QED nuclear recoil corrections for the effective potential (\ref{EffPot}) in order to take into account the screening effects.

Finally, one has to consider the nuclear polarization effects. These effects are related to the intrinsic nuclear dynamics. We have taken into account this correction for all ions using the results of Refs. \cite{Plunien:1991:5853,Plunien:1995:1119:note,Nefiodov:1996:227,Nefiodov:2002:081802,Volotka:2014:023002} and the corresponding prescriptions from Ref. \cite{Yerokhin:hydrogen:preprint}. 

We note that the numerical procedures are checked by performing the calculations of the QED corrections in two gauges, the Feynman and the Coulomb ones. A good agreement between both calculations is found. The Fermi model with a thickness parameter of $2.3$~fm is used to describe the nuclear charge distribution. The nuclear radii are taken from Refs. \cite{Angeli:2013:69}. In the cases of $Z$ with no experimental radii available, we use the root-mean-square (rms) radii obtained by employing the approximate formula from Ref.~\cite{Johnson:1985:405} and prescribe  a 1\% uncertainty to them.

\section{Numerical results and discussions \label{sec:2}}

{
\renewcommand{\arraystretch}{0.85}
\begin{table}
\caption{Individual contributions to the ionization potential of the $2s$ electron in berylliumlike calcium (in eV). 
See text for details.}\label{table:Ca}
\resizebox{\columnwidth}{!}{%
\begin{tabular}{c@{\quad}r@{\quad}r@{\quad}r@{\quad}r@{\quad}r@{\quad}r@{\quad}r}
\hline 
\hline 
\multicolumn{1}{c}{Contribution} & \multicolumn{1}{c}{CH}     & 
\multicolumn{1}{c}{LDF3}            & \multicolumn{1}{c}{KS3}    & 
\multicolumn{1}{c}{PZ3}              & \multicolumn{1}{c}{LDF4}  & 
\multicolumn{1}{c}{KS4}              & \multicolumn{1}{c}{PZ4}           \\ 
\hline 

$E^{(0)}_{\rm Dirac}$               &  $-1147.45480$  &  $-1149.96374$  &  $-1154.32591$  &  $-1157.78287$  &  $-1074.07837$  &  $-1066.78185$  &  $-1079.51335$  \\

$E^{(1)}_{\rm int}$                 &  $65.89672$     &  $68.74389$     &  $73.56661$     &  $77.27720$     &  $-7.77584$     &  $-14.87760$    &  $-2.02387$     \\

$E^{(2)}_{\rm int,Breit}$           &  $-9.77636$     &  $-9.37458$     &  $-9.52861$     &  $-12.17568$    &  $-7.48702$     &  $-7.08279$     &  $-9.28264$     \\

$E^{(2)}_{\rm int,QED}$             &  $0.00057$      &  $0.00055$      &  $0.00053$      &  $0.00054$      &  $0.00055$      &  $0.00051$      &  $0.00054$      \\

$E^{(\geqslant3)}_{\rm int,Breit}$  &  $3.88459$      &  $3.14463$      &  $2.83836$      &  $5.23177$      &  $1.89125$      &  $1.29221$      &  $3.37002$      \\

\hline
$E_{\rm int,total}$                 &  $-1087.44927$  &  $-1087.44925$  &  $-1087.44902$  &  $-1087.44903$  &  $-1087.44944$  &  $-1087.44952$  &  $-1087.44931$  \\
\hline

$E^{(1)}_{\rm SE}$                  &  $0.19895$      &  $0.19917$      &  $0.20072$      &  $0.20187$      &  $0.19207$      &  $0.18907$      &  $0.19398$      \\

$E^{(1)}_{\rm VP}$                  &  $-0.01388$     &  $-0.01388$     &  $-0.01399$     &  $-0.01408$     &  $-0.01340$     &  $-0.01320$     &  $-0.01354$     \\

$E^{(2)}_{\rm ScrSE}$               &  $-0.02115$     &  $-0.02142$     &  $-0.02314$     &  $-0.02439$     &  $-0.01397$     &  $-0.01073$     &  $-0.01600$     \\

$E^{(2)}_{\rm ScrVP}$               &  $0.00146$      &  $0.00146$      &  $0.00158$      &  $0.00168$      &  $0.00096$      &  $0.00074$      &  $0.00111$      \\

$E^{(2)}_{\rm 2loop}$               &  $-0.00016$     &  $-0.00016$     &  $-0.00016$     &  $-0.00016$     &  $-0.00016$     &  $-0.00016$     &  $-0.00016$     \\

\hline
$E_{\rm QED,total}$                 &  $0.16521$      &  $0.16516$      &  $0.16501$      &  $0.16491$      &  $0.16549$      &  $0.16573$      &  $0.16538$      \\
\hline

$E_{\rm rec,Breit}$                 &  $0.01415$      &  $0.01415$      &  $0.01415$      &  $0.01415$      &  $0.01415$      &  $0.01415$      &  $0.01415$      \\
 
$E_{\rm rec,QED}$                   &  $0.00006$      &  $0.00006$      &  $0.00006$      &  $0.00006$      &  $0.00006$      &  $0.00006$      &  $0.00006$      \\

\hline
\hline
$E_{\rm total}$                     &  $-1087.26985$  &  $-1087.26988$  &  $-1087.26980$  &  $-1087.26990$  &  $-1087.26974$  &  $-1087.26958$  &  $-1087.26972$  \\
\hline
\hline

\end{tabular}%
} 
\end{table}
}

{
\renewcommand{\arraystretch}{0.85}
\begin{table}
\caption{Individual contributions to the ionization potential of the $2s$ electron in berylliumlike xenon (in eV). 
See text for details.}\label{table:Xe}
\resizebox{\columnwidth}{!}{%
\begin{tabular}{c@{\quad}r@{\quad}r@{\quad}r@{\quad}r@{\quad}r@{\quad}r@{\quad}r}
\hline 
\hline 
\multicolumn{1}{c}{Contribution} & \multicolumn{1}{c}{CH}     & 
\multicolumn{1}{c}{LDF3}            & \multicolumn{1}{c}{KS3}    & 
\multicolumn{1}{c}{PZ3}              & \multicolumn{1}{c}{LDF4}  & 
\multicolumn{1}{c}{KS4}              & \multicolumn{1}{c}{PZ4}           \\ 
\hline 

$E^{(0)}_{\rm Dirac}$               &  $-9786.2460$  &  $-9796.3872$  &  $-9809.3421$  &  $-9821.1550$  &  $-9569.5159$  &  $-9548.0653$  &  $-9587.1179$  \\

$E^{(1)}_{\rm int}$                 &  $198.9440$    &  $209.5581$    &  $223.0184$    &  $235.1590$    &  $-18.0970$    &  $-39.3573$    &  $-0.1378$     \\

$E^{(2)}_{\rm int,Breit}$           &  $-13.6879$    &  $-13.2832$    &  $-13.4699$    &  $-16.6451$    &  $-11.1410$    &  $-10.7201$    &  $-13.2829$    \\

$E^{(2)}_{\rm int,QED}$             &  $0.0310$      &  $0.0302$      &  $0.0292$      &  $0.0298$      &  $0.0306$      &  $0.0299$      &  $0.0305$      \\

$E^{(\geqslant3)}_{\rm int,Breit}$  &  $3.9484$      &  $3.0718$      &  $2.7540$      &  $5.6013$      &  $1.7125$      &  $1.1030$      &  $3.4975$      \\

\hline
$E_{\rm int,total}$                 &  $-9597.0104$  &  $-9597.0103$  &  $-9597.0105$  &  $-9597.0101$  &  $-9597.0108$  &  $-9597.0098$  &  $-9597.0106$  \\
\hline

$E^{(1)}_{\rm SE}$                  &  $7.2387$      &  $7.2471$      &  $7.2704$      &  $7.2871$      &  $7.1549$      &  $7.1218$      &  $7.1828$      \\

$E^{(1)}_{\rm VP}$                  &  $-0.9372$     &  $-0.9382$     &  $-0.9416$     &  $-0.9438$     &  $-0.9265$     &  $-0.9229$     &  $-0.9305$     \\

$E^{(2)}_{\rm ScrSE}$               &  $-0.2890$     &  $-0.2979$     &  $-0.3221$     &  $-0.3395$     &  $-0.2039$     &  $-0.1700$     &  $-0.2327$     \\

$E^{(2)}_{\rm ScrVP}$               &  $0.0380$      &  $0.0391$      &  $0.0427$      &  $0.0450$      &  $0.0272$      &  $0.0235$      &  $0.0313$      \\

$E^{(2)}_{\rm 2loop}$               &  $-0.0165$     &  $-0.0165$     &  $-0.0165$     &  $-0.0165$     &  $-0.0165$     &  $-0.0165$     &  $-0.0165$     \\

\hline
$E_{\rm QED,total}$                 &  $6.0340$      &  $6.0336$      &  $6.0328$      &  $6.0322$      &  $6.0351$      &  $6.0358$      &  $6.0343$      \\
\hline

$E_{\rm rec,Breit}$                 &  $0.0386$      &  $0.0386$      &  $0.0386$      &  $0.0386$      &  $0.0386$      &  $0.0386$      &  $0.0386$      \\

$E_{\rm rec,QED}$                   &  $0.0031$      &  $0.0031$      &  $0.0032$      &  $0.0032$      &  $0.0031$      &  $0.0031$      &  $0.0031$      \\

$E_{\rm nucl.pol.}$                 &  $-0.0002$     &  $-0.0002$     &  $-0.0002$     &  $-0.0002$     &  $-0.0002$     &  $-0.0002$     &  $-0.0002$     \\

\hline
\hline
$E_{\rm total}$                     &  $-9590.9349$  &  $-9590.9352$  &  $-9590.9362$  &  $-9590.9363$  &  $-9590.9342$  &  $-9590.9325$  &  $-9590.9347$  \\
\hline
\hline

\end{tabular}%
} 
\end{table}
}

{
\renewcommand{\arraystretch}{0.85}
\begin{table}
\caption{Individual contributions to the ionization potential of the $2s$ electron in berylliumlike uranium (in eV). 
See text for details.}\label{table:U}
\resizebox{\columnwidth}{!}{%
\begin{tabular}{c@{\quad}r@{\quad}r@{\quad}r@{\quad}r@{\quad}r@{\quad}r@{\quad}r}
\hline 
\hline 
\multicolumn{1}{c}{Contribution} & \multicolumn{1}{c}{CH}     & 
\multicolumn{1}{c}{LDF3}            & \multicolumn{1}{c}{KS3}    & 
\multicolumn{1}{c}{PZ3}              & \multicolumn{1}{c}{LDF4}  & 
\multicolumn{1}{c}{KS4}              & \multicolumn{1}{c}{PZ4}           \\ 
\hline 

$E^{(0)}_{\rm Dirac}$               &  $-32827.813$  &  $-32858.372$  &  $-32886.232$  &  $-32913.911$  &  $-32407.665$  &  $-32369.166$  &  $-32445.215$  \\

$E^{(1)}_{\rm int}$                 &  $406.898$     &  $438.333$     &  $466.938$     &  $495.120$     &  $-13.679$     &  $-51.938$     &  $24.367$      \\

$E^{(2)}_{\rm int,Breit}$           &  $-18.134$     &  $-17.775$     &  $-18.129$     &  $-23.634$     &  $-14.308$     &  $-13.906$     &  $-17.824$     \\

$E^{(2)}_{\rm int,QED}$             &  $0.275$       &  $0.273$       &  $0.270$       &  $0.272$       &  $0.278$       &  $0.277$       &  $0.277$       \\

$E^{(\geqslant3)}_{\rm int,Breit}$  &  $6.398$       &  $5.165$       &  $4.778$       &  $9.778$       &  $2.997$       &  $2.352$       &  $6.018$       \\

\hline
$E_{\rm int,total}$                 &  $-32432.376$  &  $-32432.376$  &  $-32432.375$  &  $-32432.375$  &  $-32432.378$  &  $-32432.382$  &  $-32432.377$  \\
\hline

$E^{(1)}_{\rm SE}$                  &  $62.677$      &  $62.770$      &  $62.923$      &  $63.026$      &  $62.280$      &  $62.185$      &  $62.455$      \\

$E^{(1)}_{\rm VP}$                  &  $-14.951$     &  $-14.980$     &  $-15.029$     &  $-15.049$     &  $-14.863$     &  $-14.855$     &  $-14.910$     \\

$E^{(2)}_{\rm ScrSE}$               &  $-1.746$      &  $-1.843$      &  $-2.002$      &  $-2.109$      &  $-1.346$      &  $-1.250$      &  $-1.526$      \\

$E^{(2)}_{\rm ScrVP}$               &  $0.423$       &  $0.454$       &  $0.506$       &  $0.526$       &  $0.335$       &  $0.329$       &  $0.383$       \\

$E^{(2)}_{\rm 2loop}$               &  $-0.244$      &  $-0.244$      &  $-0.244$      &  $-0.244$      &  $-0.244$      &  $-0.244$      &  $-0.244$      \\

\hline
$E_{\rm QED,total}$                 &  $46.158$      &  $46.156$      &  $46.154$      &  $46.150$      &  $46.162$      &  $46.164$      &  $46.158$      \\
\hline

$E_{\rm rec,Breit}$                 &  $0.066$       &  $0.066$       &  $0.066$       &  $0.066$       &  $0.066$       &  $0.066$       &  $0.066$       \\

$E_{\rm rec,QED}$                   &  $0.047$       &  $0.047$       &  $0.047$       &  $0.047$       &  $0.046$       &  $0.046$       &  $0.046$       \\

$E_{\rm nucl.pol.}$                 &  $-0.036$      &  $-0.036$      &  $-0.036$      &  $-0.036$      &  $-0.036$      &  $-0.036$      &  $-0.036$      \\

\hline
\hline
$E_{\rm total}$                     &  $-32386.141$  &  $-32386.142$  &  $-32386.144$  &  $-32386.148$  &  $-32386.139$  &  $-32386.141$  &  $-32386.142$  \\
\hline
\hline

\end{tabular}%
}
\end{table}
}

The individual contributions to the $2s$ electron ionization potentials of berylliumlike calcium, xenon, and uranium evaluated for various screening potentials are presented in Tables \ref{table:Ca}--\ref{table:U}, respectively. 
The first line contains the ionization potentials obtained by solving Eq.~(\ref{DirEq}) for the corresponding effective potential with the finite nuclear size effect included, $E^{(0)}_{\rm Dirac} = \varepsilon_{2s} - mc^2$. 
The nuclear deformation correction was added  for uranium ion in accordance with the results of Ref.~\cite{Kozhedub:2008:032501}. 
$E^{(1)}_{\rm int}$  stands for the contribution of the first-order diagrams in Fig.~\ref{fig:int} [diagrams (a) and (e)]. The diagram (a) was evaluated keeping the energy dependence of the photon propagator in the exchange part, i.e., in the framework of QED. 
The contribution of the second-order diagrams evaluated in the Breit approximation is given in the third line. As it was mentioned in Sec.~\ref{sec:1}, here the Breit approximation implies the calculations performed in the Coulomb gauge at zero energy transfer and neglecting the contribution of the negative-energy continuum. We note that this approach to the Breit approximation differs, e.g., from the one used by Yerokhin \textit{et al.} in Ref.~\cite{Yerokhin:2007:062501}, where  the exchange by two Breit photons was considered as belonging to the higher order corrections. Furthermore, the influence of the negative energy continuum was partly accounted for in Ref.~\cite{Yerokhin:2007:062501} in the contribution considered. 
In the fourth row we give the QED correction $E^{(2)}_{\rm int,QED}$ to the third line, $E^{(2)}_{\rm int,Breit}$. It is calculated as the difference between the contribution of the second-order diagrams in Fig.~\ref{fig:int}  calculated within the rigorous QED approach and within the Breit approximation. 
The interelectronic-interaction corrections $E^{(\geqslant3)}_{\rm int,Breit}$ of the third and higher orders evaluated with the CI-DFS method are shown in the fifth row. In the sixth line the sum $E_{\rm int,total}$ of all the previous terms (the rows from first to fifth) is presented. From Tables \ref{table:Ca}--\ref{table:U} it can be seen that for all seven screening potentials the $E_{\rm int,total}$ values are in a good agreement with each other. 
The next four rows contain the contributions of the QED diagrams from Fig.~\ref{fig:se}. 
The lines labeled as $E^{(1)}_{\rm SE}$ and $E^{(1)}_{\rm VP}$ correspond to the first-order SE and VP corrections, whereas the lines $E^{(2)}_{\rm ScrSE}$ and $E^{(2)}_{\rm ScrVP}$ present the related SE and VP two-electron corrections.
The two-loop one-electron corrections for the $2s$ electron are collected together in the eleventh line, $E^{(2)}_{\rm 2loop}$.  
The sum of all QED contributions (the rows from seventh to eleventh) is given in the line labeled with $E_{\rm QED,total}$. Again, it is seen that the results of the calculations are in a good agreement for all screening potentials. One can check that this holds independently for the SE and VP corrections. 
The next two rows contain the nuclear recoil contribution evaluated within the Breit approximation and the QED recoil correction, respectively. For uranium  and xenon  we also give  the contribution of the nuclear polarization effect in the line $E_{\rm nucl.pol.}$. 
Finally, in the last line we present the total results for the $2s$ electron ionization energies. From Tables~\ref{table:Ca}--\ref{table:U} it can be seen, that the total values of the ionization potentials are almost independent of the type of the screening potential used in the calculations. 
For this reason, for all other ions in the range $16 \leqslant Z \leqslant 96$ the calculations have been performed for  two screening potentials only, namely for $V_{\rm LDF3}$ and $V_{\rm LDF4}$. We have chosen these two potentials in order to control the accuracy of the calculations along the isoelectronic sequence.

Table~\ref{table:IP} displays the ionization potentials for all berylliumlike ions with $Z=16-96$. For calcium, xenon and uranium we give the averages of the total values evaluated for the seven screening potentials. For the other ions the ionization potentials were obtained by averaging the results of calculations with the LDF3 and LDF4 potentials. The uncertainties given in the parentheses were obtained by summing quadratically the uncertainty due to the nuclear size effect, the uncertainty of the CI-DFS calculation, and the uncertainties due to uncalculated higher-order QED contributions.  For uranium ion the nuclear size uncertainty was estimated in accordance with Ref.~\cite{Kozhedub:2008:032501} taking into account the nuclear deformation correction. 
For the other ions we estimated this uncertainty by adding quadratically two contributions. The first one was calculated by varying the rms nuclear radius within its error bar. The second one rather conservatively estimates the uncertainty in the nuclear charge distribution. It was obtained by varying the distribution model from the Fermi model to the homogeneously charged sphere model. 
The uncertainty due to uncalculated QED corrections to the interelectronic-interaction contributions  can be conservatively estimated as the product of the term $E^{(\geqslant3)}_{\rm int,Breit}$  and the doubled ratio of the second-order interelectronic-interaction QED correction to the  corresponding contribution calculated within the Breit approximation, $E^{(2)}_{\rm int,QED}/E^{(2)}_{\rm int,Breit}$. The contribution of the higher-order screened QED diagrams can be estimated by multiplying the second-order QED term by the factor $2/Z$. 
On the other hand, the total results should be fully independent of the zeroth-order approximation potential, provided the interelectronic-interaction and QED corrections are calculated to all orders. Therefore, the discrepancy between the calculations with different potentials may serve as an estimation of the uncalculated QED contributions. We have used both estimations of the uncalculated QED corrections and summed them quadratically. 
Finally, we have estimated the uncertainty due to uncalculated two-loop one-electron QED contributions according to Ref.~\cite{Yerokhin:2008:062510}. 
For low-$Z$ ions the main source of the uncertainty is the CI-DFS calculation. For high-$Z$ ions the   nuclear size uncertainty and the uncertainties due to uncalculated QED corrections start to play an important role.

In Table~\ref{table:IP} we compare our results for the ionization potentials of berylliumlike ions with  theoretical predictions by other authors. It can be seen that, as a rule, our ionization potentials are in a good agreement with the results of the previous calculations, but have much higher accuracy. This is not surprising since, as was noted in Sec.~\ref{sec:0}, in all previous evaluations of the ionization potentials of Be-like ions the QED effects were included either semiempirically or within some one-electron approximations.  In our previous work~\cite{Malyshev:2014:062517} we tabulated the ionization potentials for even-$Z$ ions in the range $18\leqslant Z\leqslant 96$. The ionization potentials  were obtained as the differences  between the binding energies of Be-like ions calculated directly in Ref.~\cite{Malyshev:2014:062517} and the binding energies of Li-like ions compiled from the previous tabulations. In the present work the ionization potentials were calculated directly and self-consistently. This resulted in a three-times improvement of the theoretical precision for heavy ions and in a two-times improvement for low- and middle-$Z$ ions compared to our previous calculations \cite{Malyshev:2014:062517}.

{
\renewcommand{\arraystretch}{0.85}
\begin{table}
\caption{Ionization potentials (in eV) for berylliumlike ions with $Z = 16-96$.}\label{table:IP}
\begin{center}
\begin{tabular}{l@{\qquad\qquad\qquad}l@{\qquad\qquad\qquad}l}
 \hline
 \hline
Nucleus & This work & Other works \\
 \hline

   $^{32}_{16}$S   &   $-652.135(20)$         &      $-651.96(12)^{\rm a}$      \\
                                &             &      $-652.20^{\rm b}$            \\
                                &             &      $-652.1923^{\rm c}$        \\
                                &             &      $-652.3391^{\rm d}$        \\                               
  $^{35}_{17}$Cl   &   $-750.465(20)$         &      $-750.23(19)^{\rm a}$      \\
                                &             &      $-750.54^{\rm b}$             \\
                                &             &      $-750.5212^{\rm c}$        \\
                                &             &      $-750.7090^{\rm d}$        \\                               
  $^{40}_{18}$Ar   &   $-855.750(20)$         &      $-855.47(27)^{\rm a}$      \\
                                &             &      $-855.82^{\rm b}$             \\
                                &             &      $-855.8045^{\rm c}$         \\
                                &             &      $-856.0432^{\rm d}$         \\                               
  $^{39}_{19}$K    &   $-968.010(20)$         &      $-967.66(35)^{\rm a}$      \\
                                &             &      $-968.08^{\rm b}$             \\
                                &             &      $-968.0597^{\rm c}$         \\                            
  $^{40}_{20}$Ca  &   $-1087.270(20)$         &      $-1086.85(40)^{\rm a}$      \\
                                &             &      $-1087.3^{\rm b}$               \\
                                &             &      $-1087.311^{\rm c}$          \\        
                                &             &      $-1087.44^{\rm e}$           \\                                                         
  $^{45}_{21}$Sc  &   $-1213.555(20)$         &      $-1213.06(45)^{\rm a}$      \\
                                &             &      $-1213.6^{\rm b}$               \\
                                &             &      $-1213.583^{\rm c}$           \\                            
  $^{48}_{22}$Ti   &   $-1346.889(20)$        &      $-1346.33(47)^{\rm a}$      \\
                                &             &      $-1347.0^{\rm b}$              \\
                                &             &      $-1346.899^{\rm c}$          \\                            
  $^{51}_{23}$V   &   $-1487.301(20)$         &      $-1486.67(52)^{\rm a}$      \\
                                &             &      $-1487.4^{\rm b}$              \\
                                &             &      $-1487.285^{\rm c}$          \\                            
  $^{52}_{24}$Cr  &   $-1634.820(20)$         &      $-1634.11(55)^{\rm a}$      \\
                                &             &      $-1634.9^{\rm b}$              \\
                                &             &      $-1634.769^{\rm c}$          \\                            
  $^{55}_{25}$Mn  &   $-1789.478(20)$         &      $-1788.70(62)^{\rm a}$      \\
                                &             &      $-1789.6^{\rm b}$               \\
                                &             &      $-1789.381^{\rm c}$            \\                            

\end{tabular}
\end{center}
\end{table}
}
\addtocounter{table}{-1}
{
\renewcommand{\arraystretch}{0.85}
\begin{table}
\caption{(Continued.)}
\begin{center}
\begin{tabular}{l@{\qquad\qquad\qquad}l@{\qquad\qquad\qquad}l}
 \hline
 \hline
Nucleus & This work & Other works \\
 \hline
 
  $^{56}_{26}$Fe   &   $-1951.307(21)$        &      $-1950.4(1.8)^{\rm a}$      \\
                                &             &      $-1951.4^{\rm b}$              \\
                                &             &      $-1951.10^{\rm f}$            \\                            
  $^{59}_{27}$Co  &   $-2120.343(22)$         &      $-2119.4(1.9)^{\rm a}$      \\
                                &             &      $-2120.5^{\rm b}$             \\
                                &             &      $-2120.14^{\rm f}$            \\                            
  $^{58}_{28}$Ni  &   $-2296.621(23)$         &      $-2295.6(2.1)^{\rm a}$      \\
                                &             &      $-2296.7^{\rm b}$             \\
                                &             &      $-2296.42^{\rm f}$            \\      
                                &             &      $-2298.27^{\rm e}$           \\                                                         
  $^{63}_{29}$Cu  &   $-2480.182(23)$         &      $-2479.1(2.2)^{\rm a}$      \\
                                &             &      $-2480.3^{\rm b}$              \\
                                &             &      $-2479.98^{\rm f}$            \\                            
 $^{64}_{30}$Zn   &   $-2671.064(24)$         &      $-2669.9(2.5)^{\rm a}$      \\
                                &             &      $-2671.2^{\rm b}$              \\
                                &             &      $-2670.84^{\rm f}$             \\                            
                                &             &      $-2673.10^{\rm e}$            \\ 
 $^{69}_{31}$Ga   &   $-2869.311(25)$         &      $-2868.1(2.8)^{\rm a}$      \\
                                &             &      $-2869.5^{\rm b}$              \\
                                &             &      $-2869.08^{\rm f}$             \\                            
 $^{74}_{32}$Ge   &   $-3074.967(25)$         &      $-3073.6(3.0)^{\rm a}$      \\
                                &             &      $-3075.1^{\rm b}$              \\
                                &             &      $-3074.73^{\rm f}$             \\                            
 $^{75}_{33}$As   &   $-3288.077(26)$         &      $-3286.6(3.2)^{\rm a}$      \\
                                &             &      $-3288.2^{\rm b}$             \\
                                &             &      $-3287.84^{\rm f}$            \\                            
 $^{80}_{34}$Se   &   $-3508.691(27)$         &      $-3507.1(3.5)^{\rm a}$      \\
                                &             &      $-3508.9^{\rm b}$             \\
                                &             &      $-3508.45^{\rm f}$            \\                            
$^{79}_{35}$Br    &   $-3736.857(27)$         &      $-3735.3(3.7)^{\rm a}$      \\
                                &             &      $-3737.0^{\rm b}$             \\
                                &             &      $-3736.65^{\rm f}$            \\                                          

\end{tabular}
\end{center}
\end{table}
}
\addtocounter{table}{-1}
{
\renewcommand{\arraystretch}{0.85}
\begin{table}
\caption{(Continued.)}
\begin{center}
\begin{tabular}{l@{\qquad\qquad\qquad}l@{\qquad\qquad\qquad}l}
 \hline
 \hline
Nucleus & This work & Other works \\
 \hline
          
 $^{84}_{36}$Kr   &   $-3972.631(28)$         &      $-3970.8(4.0)^{\rm a}$      \\
                                &             &      $-3972.8^{\rm b}$             \\
                                &             &      $-3972.45^{\rm f}$           \\   
                                &             &      $-3975.7297^{\rm d}$       \\                                      
$^{85}_{37}$Rb   &   $-4216.065(28)$          &      $-4214.2(4.2)^{\rm a}$      \\
                                &             &      $-4216.3^{\rm b}$             \\                         
 $^{88}_{38}$Sr   &   $-4467.217(29)$         &      $-4465.3(4.5)^{\rm a}$      \\
                                &             &      $-4467.4^{\rm b}$            \\                         
$^{89}_{39}$Y     &   $-4726.144(30)$         &      $-4724.0(4.7)^{\rm a}$      \\
                                &             &      $-4726.4^{\rm b}$            \\                         
 $^{90}_{40}$Zr   &   $-4992.909(30)$         &      $-4990.7(5.0)^{\rm a}$      \\
                                &             &      $-4993.1^{\rm b}$             \\                         
 $^{93}_{41}$Nb   &   $-5267.575(31)$         &      $-5265.2(5.2)^{\rm a}$      \\
                                &             &      $-5267.8^{\rm b}$              \\                         
 $^{98}_{42}$Mo   &   $-5550.207(31)$         &      $-5547.8(5.5)^{\rm a}$      \\
                                 &            &      $-5550.5^{\rm b}$              \\                         
$^{97}_{43}$Tc     &   $-5840.876(32)$        &      $-5838.3(5.6)^{\rm a}$      \\
                                &             &      $-5841.1^{\rm b}$              \\                         
 $^{102}_{44}$Ru  &   $-6139.651(33)$         &      $-6136.8(5.8)^{\rm a}$      \\
                                &             &      $-6139.9^{\rm b}$               \\                         
 $^{103}_{45}$Rh  &   $-6446.608(33)$         &      $-6443.7(6.1)^{\rm a}$      \\
                                &             &      $-6446.9^{\rm b}$              \\                         
 $^{106}_{46}$Pd  &   $-6761.822(34)$         &      $-6758.8(6.3)^{\rm a}$      \\
                                &             &      $-6762.1^{\rm b}$               \\                         
 $^{107}_{47}$Ag  &   $-7085.373(34)$         &      $-7082.1(6.6)^{\rm a}$      \\
                                &             &      $-7085.7^{\rm b}$               \\                         
 $^{114}_{48}$Cd  &   $-7417.341(34)$         &      $-7413.9(6.8)^{\rm a}$      \\
                                &             &      $-7417.7^{\rm b}$               \\                         
 $^{115}_{49}$In  &   $-7757.815(34)$         &      $-7754.2(7.1)^{\rm a}$      \\
                                &             &      $-7758.2^{\rm b}$              \\                         
 $^{120}_{50}$Sn  &   $-8106.879(35)$         &      $-8103.1(7.3)^{\rm a}$      \\
                                &             &      $-8107.2^{\rm b}$              \\                         
 $^{121}_{51}$Sb  &   $-8464.627(35)$         &      $-8455(4)^{\rm g}$      \\
                                &             &      $-8465.0^{\rm b}$         \\                                  

\end{tabular}
\end{center}
\end{table}
}
\addtocounter{table}{-1}
{
\renewcommand{\arraystretch}{0.85}
\begin{table}
\caption{(Continued.)}
\begin{center}
\begin{tabular}{l@{\qquad\qquad\qquad}l@{\qquad\qquad\qquad}l}
 \hline
 \hline
Nucleus & This work & Other works \\
 \hline
      
$^{130}_{52}$Te   &   $-8831.151(35)$         &      $-8821(4)^{\rm g}$      \\
                                &             &      $-8831.5^{\rm b}$         \\                                
$^{127}_{53}$I     &   $-9206.555(36)$        &      $-9196(4)^{\rm g}$      \\
                                &             &      $-9207.0^{\rm b}$        \\                         
$^{132}_{54}$Xe   &   $-9590.935(37)$         &      $-9581(4)^{\rm g}$      \\
                                &             &      $-9591.4^{\rm b}$        \\                          
$^{133}_{55}$Cs   &   $-9984.401(37)$         &      $-9974(4)^{\rm g}$      \\
                                &             &      $-9984.9^{\rm b}$        \\                          
                                &             &      $-10002.68^{\rm e}$      \\  
$^{138}_{56}$Ba   &   $-10387.058(38)$        &      $-10376(4)^{\rm g}$      \\
                                &             &      $-10388^{\rm b}$            \\                         
$^{139}_{57}$La   &   $-10799.024(39)$        &      $-10789(5)^{\rm g}$      \\
                                &             &      $-10800^{\rm b}$           \\                         
$^{140}_{58}$Ce   &   $-11220.414(40)$        &      $-11210(5)^{\rm g}$      \\
                                &             &      $-11221^{\rm b}$            \\                           
$^{141}_{59}$Pr   &   $-11651.351(41)$        &      $-11641(5)^{\rm g}$      \\
                                &             &      $-11652^{\rm b}$           \\                           
$^{142}_{60}$Nd   &   $-12091.960(42)$        &      $-12082(5)^{\rm g}$      \\
                                &             &      $-12092^{\rm b}$            \\                           
$^{145}_{61}$Pm   &   $-12542.365(47)$        &      $-12532(6)^{\rm g}$      \\                         
$^{152}_{62}$Sm   &   $-13002.682(45)$        &      $-12992(6)^{\rm g}$      \\                         
$^{153}_{63}$Eu   &   $-13473.107(46)$        &      $-13462(7)^{\rm g}$      \\                         
$^{158}_{64}$Gd   &   $-13953.751(48)$        &      $-13943(10)^{\rm g}$      \\                         
$^{159}_{65}$Tb   &   $-14444.844(97)$        &      $-14434(12)^{\rm g}$      \\                         
$^{164}_{66}$Dy   &   $-14946.339(52)$        &      $-14936(15)^{\rm g}$      \\                         
$^{165}_{67}$Ho   &   $-15458.633(58)$        &      $-15448(19)^{\rm g}$      \\                         
$^{166}_{68}$Er   &   $-15981.752(56)$        &      $-15971(24)^{\rm g}$      \\                         
$^{169}_{69}$Tm   &   $-16515.973(59)$        &      $-16505(30)^{\rm g}$      \\                         
$^{174}_{70}$Yb    &   $-17061.335(61)$       &      $-17050(40)^{\rm g}$      \\                         
$^{175}_{71}$Lu    &   $-17618.130(71)$       &      $-17607(40)^{\rm g}$      \\                         
$^{180}_{72}$Hf    &   $-18186.626(67)$       &      $-18176(50)^{\rm g}$      \\                         
$^{181}_{73}$Ta    &   $-18766.923(70)$       &      $-18756(50)^{\rm g}$      \\                         
$^{184}_{74}$W    &   $-19359.249(74)$        &      $-19362.5(3.1)^{\rm h}$      \\       
                                  &           &      $-19348(50)^{\rm g}$      \\                                                 

\end{tabular}
\end{center}
\end{table}
}
\addtocounter{table}{-1}
{
\renewcommand{\arraystretch}{0.85}
\begin{table}
\caption{(Continued.)}
\begin{center}
\begin{tabular}{l@{\qquad\qquad\qquad}l@{\qquad\qquad\qquad}l}
 \hline
 \hline
Nucleus & This work & Other works \\
 \hline
    
$^{187}_{75}$Re    &   $-19963.856(81)$      &      $-19953(60)^{\rm g}$      \\              
$^{192}_{76}$Os    &   $-20580.902(81)$      &      $-20570(70)^{\rm g}$      \\                              
$^{193}_{77}$Ir    &   $-21210.77(22)$       &      $-21200(90)^{\rm g}$      \\                   
$^{194}_{78}$Pt    &   $-21853.579(89)$      &      $-21843(90)^{\rm g}$      \\                   
$^{197}_{79}$Au    &   $-22509.658(95)$      &      $-22498(100)^{\rm g}$      \\                   
$^{202}_{80}$Hg    &   $-23179.23(10)$       &      $-23168(110)^{\rm g}$      \\                   
$^{205}_{81}$Tl    &   $-23862.66(10)$       &      $-23852(120)^{\rm g}$      \\                   
$^{208}_{82}$Pb    &   $-24560.15(11)$       &      $-24548(120)^{\rm g}$      \\                   
$^{209}_{83}$Bi    &   $-25272.10(12)$       &      $-25260(140)^{\rm g}$      \\                   
$^{210}_{84}$Po    &   $-25998.67(14)$       &      $-25988(150)^{\rm g}$      \\                   
$^{215}_{85}$At    &   $-26740.49(28)$       &      $-26729(160)^{\rm g}$      \\                   
$^{220}_{86}$Rn    &   $-27497.35(17)$       &      $-27486(170)^{\rm g}$      \\                   
$^{223}_{87}$Fr    &   $-28270.33(18)$       &      $-28259(190)^{\rm g}$      \\                   
$^{226}_{88}$Ra    &   $-29059.47(24)$       &      $-29048(200)^{\rm g}$      \\                   
$^{227}_{89}$Ac    &   $-29865.71(41)$       &      $-29854(220)^{\rm g}$      \\                   
$^{232}_{90}$Th    &   $-30687.91(20)$       &      $-30677(240)^{\rm g}$      \\                   
$^{231}_{91}$Pa    &   $-31528.98(50)$       &      $-31516(250)^{\rm g}$      \\                   
$^{238}_{92}$U     &   $-32386.14(20)$       &      $-32374(300)^{\rm g}$      \\                   
$^{237}_{93}$Np    &   $-33263.97(61)$       &      $-33252(300)^{\rm g}$      \\                   
$^{240}_{94}$Pu    &   $-34158.55(48)$       &      $-34147(300)^{\rm g}$      \\                   
$^{243}_{95}$Am    &   $-35073.44(27)$       &      $-35062(300)^{\rm g}$      \\                   
$^{244}_{96}$Cm    &   $-36009.57(37)$       &      $-35996(400)^{\rm g}$      \\                   

\hline
\end{tabular}
\end{center}
\bigskip
\raggedright

$^{\rm a}$ Bi\'emont \textit{et al.} \cite{Biemont:1999:117}   

$^{\rm b}$ Gu \cite{Gu:2005:267}.

$^{\rm c}$ Chung \textit{et al.} \cite{Chung:1993:1740}.

$^{\rm d}$ Pathak \textit{et al.} \cite{Pathak:2014:042510}.

$^{\rm e}$ Chaudhuri \textit{et al.} \cite{Chaudhuri:2000:5129}.

$^{\rm f}$ Yong-Qiang \textit{et al.} \cite{Yong-Qiang:2008:3627}

$^{\rm g}$ Rodrigues \textit{et al.} \cite{Rodrigues:2004:117} with the uncertainty 
prescribed by NIST \cite{NIST:2014}.

$^{\rm h}$ Kramida and Reader \cite{Kramida:2006:457}.

\end{table}
}

\section{Summary \label{sec:3}}

To summarize, we have performed the \textit{ab initio} QED calculations of the ground-state ionization energies for all berylliumlike ions in the range $16 \leqslant Z \leqslant 96$. Our numerical approach merges the rigorous QED evaluations in the first and second orders of the perturbation theory with the calculations of the higher-order electron-correlation contributions within the large-scale CI-DFS method. As the result, we have obtained the most precise theoretical predictions for the ionization potentials in Be-like ions. The achieved accuracy allows to probe the QED corrections in the ionization energies of berylliumlike ions and thus provides a prerequisite for tests of QED at strong fields. 
In the future, we plan to apply this approach for calculations of the transition energies in berylliumlike ions that are of current experimental interest \cite{Beiersdorfer:1998:1944,Draganic:2003:183001,Bernhardt:2015:144008}.

\section*{Acknowledgements}
We thank Vladimir Yerokhin for valuable discussions.
This work was supported by RFBR (Grants No. 13-02-00630, No. 15-03-07644, 
No. 14-02-31316, No. 14-02-00241, and No. 14-02-31476), by SPbSU
(Grants No. 11.38.269.2014, No. 11.38.261.2014, and No. 11.38.237.2015), and by  DFG (Grant No. VO 1707/1-2).
A.V.M. acknowledges the support from the Dynasty foundation and DAAD.
The work was carried out with the financial support of the FAIR-Russia
Research Center.


%

\end{document}